\documentstyle[paspconf,epsf,12pt]{article}

\begin{document}

\title{Polarized Foreground from Thermal Dust Emission}

\author{S. Prunet, A. Lazarian}

\affil{CITA, Mc Lennan Labs, 60 St George Street, Toronto ON M5S3H8,
  Canada}

\begin{abstract}
In this review, we intend to present the current knowledge of the
polarized emission from thermal dust in our Galaxy. We show different
methods to estimate the spatial distribution statistics of this emission in the
lack of any data from the diffuse ISM, and compare it to the expected
CMB polarized signal. We finally show how this contaminant could 
be efficiently removed from CMB maps using multi-frequency observations.
\end{abstract}

\section{Introduction}

The polarization of the cosmic microwave background was the subject 
of extensive theoretical studies in the last three years 
(Seljak 1997, Seljak \& Zaldarriaga 1997, Zaldarriaga \& Seljak 1997, 
Kamionkowski {\it et al.} 1997a, Kamionkowski {\it et al.} 1997b, 
Kamionkowski \& Kosowsky 1998).
Indeed, it should provide us with additional information on the
cosmological parameters, in particular the reionization optical depth
(Zaldarriaga 1997) and
the tensor-related parameters $n_T$ (spectral index of tensor perturbations), 
and $T/S$ (tensor to scalar amplitudes ratio, see
Kamionkowski {\it et al.} 1997b, Seljak \& Zaldarriaga 1997). 

Although the existence of 
polarization in the CMB has first been predicted long ago in the
context of an anisotropic universe (Rees 1968), there are so far
only upper limits on the polarization level of the cosmic microwave
background (Netterfield {\em et al.} 1995), which in the case of an 
isotropic universe is expected to be at best $10\%$ of the anisotropy level 
(Bond \& Efstathiou 1984). 
The late revival of
interest in the polarized part of the CMB is motivated by the fact
that the sensitivities of future satellite missions like MAP and {\sc
planck} are
comparable to the expected level of polarization in the CMB. 

While this is enough 
to raise the interest of theoreticians in the matter, the issue of the 
{\it measurability} of such a signal in a given instrumental
configuration needs further investigation. In particular, one needs to 
study the level of contamination of this polarized signal by other
astrophysical sources of emission as a function of  both frequency {\it and}
spatial scale. This paper intends to give the present status of our
knowledge concerning a Galactic foreground polarized emission: 
{\it the polarized
emission coming from the thermal emission of aligned
dust grains.} 

The polarization of the stellar light in {\it absorption} has been
discovered a long time ago (Hiltner 1949, Hall 1949). 
This polarization 
has been interpreted as a selective absorption of the stellar light by 
dust grains aligned respectively to magnetic field. Currently UV, optical,
near-infrared and far-infrared polarimetry is one of the major sources
of information about magnetic field structure.

The very same polarization is the impediment for the cosmic microwave
studies, as it shall contribute to the polarized signal to be measured
by MAP and PLANK. In this review we estimate the level of contamination
of polarized signal by galactic dust.

In what follows we estimate the intrinsic polarized emissivity of dust
at submillimeter wavelengths (section~2), discuss the angular distribution
of polarized dust emission (section~3) and compare the dust polarized
spectrum with that of CMB (section~4). We discuss intended work in 
section~5 and
summarize our results in section~6.

\section{Intrinsic polarized emissivity of Galactic dust at submillimeter 
wavelengths}

\subsection{Constraints on the grains shape and sizes}
\label{sec:shapes}

As we will see in this section, the polarized emissivity of 
the grains depends on their shape, more specifically
on their asphericity. There have been therefore several attempts
to constrain their shape using optical (Kim \& Martin 1995 and references
therein) and infrared data (Hildebrand \& Dragovan 1995, Lee \& Draine 1985 
and references therein). The idea is to constrain the shape of the grains
by fitting the observed polarization (in absorption) to extinction ratios 
of stellar light in some frequency range. These ratios, in the UV and 
optical wavelengths,
are very sensitive to the size distribution of the grains (because the
size of the grains and the wavelengths observed are comparable), and therefore
are mainly used to constrain it, but they nevertheless tend to show that,
assuming spheroidal shape for the grains, oblate grains are somewhat preferred 
to prolate grains. 

On the other hand, given the dielectric properties of the grain materials
in the infrared, it is possible to constrain the shape of the grains using
the same ratios, but in a frequency range covering some absorption features.
This is what has been done around the $3.1 \mu{\rm m}$ 
$H_2O$ ice feature, and around the $9.7 \mu{\rm m}$ silicate feature 
(see Hildebrand \& Dragovan 1995, Lee \& Draine 1985). 
At those wavelengths, the polarization curves are less sensitive to the
size distribution of the grains, provided that the maximal size of the
distribution remains much smaller than the wavelengths of interest.
Indeed, we are then in the conditions where the dipole approximation for 
the computation of polarized cross-sections is valid ({\em cf} Draine \& 
Lee 1984). 

Hildebrand \& Dragovan (1995), analyzing the polarization in
absorption around the $9.7 \mu{\rm m}$ silicate feature, found that this
curve was best fitted by oblate grains with an axis-ratio of $(2:3)$; this
is in agreement with the other studies of Kim \& Martin (1995) and 
Lee \& Draine (1985) who both inferred that the correct shape of the grains 
should be oblate, with an axis ratio of $(1:2)$.
This axis ratio, inferred from near-infrared and optical 
spectro-polarimetry, predicts a polarization degree in emission of
$35\%$ for perfectly aligned grains on a uniform magnetic field in
the plane of the sky. Of course, such an idealized situation is not
likely to occur, and different depolarization mechanisms
will diminish the {\em observed} degree of polarization
in the FIR wavelengths. Such mechanisms are discussed in the next section.

One could argue that the grains responsible for the near-infrared and 
optical dichroic absorption could be different from those responsible
for the FIR polarized emission (see Goodman 1995). This indeed may be
true for molecular clouds, where cold grains far from
emission sources may not be aligned (Lazarian, Goodman \& Myers 1997,
Hildebrand {\em et al.} 1999). In diffuse medium, however,
there are both theoretical and 
observational reasons to believe that adsorbing and emitting
grains are indeed the same. 

First, suppose that a grain population emits a polarized thermal radiation 
in the FIR, if we want this population not to contribute to the polarization
by absorption in the optical or near-infrared region, the size of those 
grains should be very big ($\ga 10\mu{\rm m}$), and they would then induce
a lot of absorption in the near-infrared, which is not seen (unless we grossly
overestimate the absorption properties of the grain materials). 
Another reason why such hypothesis does not seem plausible is the
measured metallicity of the diffuse interstellar medium, indeed the 
interstellar abundance of Si provides us with barely enough grain material
to build the grain mass distributions necessary to fit the optical data
(Kim {\em et al.} 1994, Grevesse \& Anders 1989).

There are also observational reasons to believe that the same grains are
responsible for the polarization by absorption in the near infra-red and
the polarized emission in the FIR. Indeed, one of the simplest tests of this
ansatz is the expected orthogonality of the direction of polarization
between the two wavelengths regimes (see below). This orthogonality was 
verified by Hildebrand {\em et al.} 1984 in Orion (see also Hildebrand 1988).
A good match of
the expected maximum polarization in emission predicted from the observed
polarization to extinction ratio in the near-infrared also supports
the idea that the grains are the same.

Indeed, if we consider an optically thin medium, we can show that the 
polarization degree in emission $P_{em}$ and polarization degree in
absorption $ P_{abs}$ are related via a simple relation
\begin{equation}
P_{em}(\lambda) = - P_{abs}(\lambda)
/\tau(\lambda)
\end{equation}
(see Hildebrand 1988). 
From the {\em observed} polarization to extinction ratio at 
$\lambda_2=2.2\mu{\rm m}$ (Jones {\em et al.} 1992) and 
using the dielectric functions of Draine \& Lee (1984) for a mixture of
``astronomical silicate'' and graphite grains to compute $P_{em}(\lambda_1)/
P_{em}(\lambda_2)$ 
Hildebrand \& Dragovan (1995) obtained, using eq.(1),
the maximum polarization one could expect at $\lambda_1=100\mu{\rm m}$, 
They find a maximum polarization of the order of $9\%$,
which corresponds to the maximum polarization observed by Hildebrand 
{\em et al.} (1995) in the Orion nebulae. 

It seems then that the same grains are responsible for the polarization of
starlight in the near-infrared and for the FIR polarized emission. We will
now expose the different depolarization mechanisms responsible for the 
discrepancy between the observed maximum polarization ($9\%$) and the 
theoretical prediction of $35 \%$.

\subsection{Polarized emissivity of the grains - Sources of depolarisation}
\label{sec:emissivity}
As shown in Greenberg (1968), the dichroism of the medium (defined as 
the {\em observed} polarization cross-section can be written as:
\begin{equation}
C_x - C_y = C_{pol}RF\cos^2(\zeta)
\end{equation}
where $(x,y)$ defines the plane of the sky, the regular magnetic field 
being in the $(y,z)$ plane, making an angle $\zeta$ with the $y$ axis (see
Fig...).
The radiation is experiencing different depolarisation effects, as it
travels towards the observer:
\begin{itemize}
\item[$\bullet$] $R = 3/2(\langle \cos^2(\beta) \rangle -1/3)$ is the
so-called ``Rayleigh reduction factor'', which gives the amount of 
depolarisation arising from the imperfect alignment of dust grains 
on the magnetic field lines, $\beta$ being the precession angle of
the grain (for a review on dust grain alignment mechanisms,
see e.g. Lazarian {\em et al.} 1997).
\item[$\bullet$] $F = 3/2(\langle \cos^2(\theta) \rangle -1/3)$ is the
reduction factor coming from a turbulent random component of the Galactic
magnetic field. In this case, $\theta$ is the angle between the random
and the regular component of the field.
\item[$\bullet$] Finally, the $\cos^2(\zeta)$ factor arises from
a projection effect.
\end{itemize}

\begin{figure}
\plotfiddle{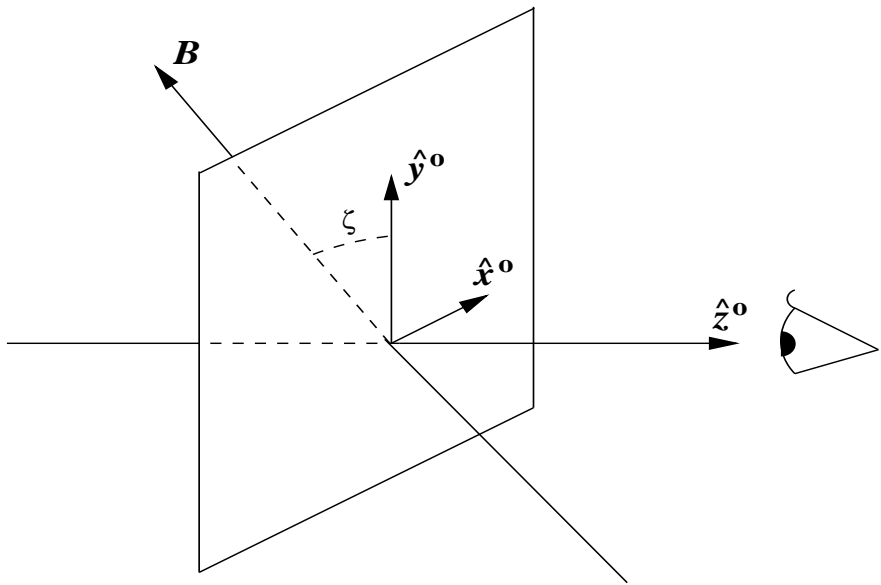}{8cm}{0}{70}{60}{-250}{-135}
\caption{The geometry of observations. Angle $\zeta$ is between the
magnetic field and the $XY$-plane. After Roberge \& Lazarian 1999.}
\end{figure}

As stressed by Goodman (1996), the observed polarization flux in absorption
depends on the direction of the magnetic field, and its fluctuations, while
the extinction does not\footnote{There is in fact a small difference of 
extinction for a population of perfectly aligned grains and for another 
population of the same grains but unaligned. This difference could account
for an error on the determination of $\tau$ no bigger than $10\%$, which
is less than the usual observational errors on $P_{abs}/\tau$ anyway
(see Lee \& Draine 1985).}. Thus, there is a large scatter in the observed
values of $P_{abs}/\tau$, which implies a scatter in the corresponding 
$P_{em}$ in the far infrared (for an illustration see Fig.~5a of Hildebrand 
1996). 

The {\em maximum observed value} of $P_{em}$ (around $10\%$ in Orion) has 
been interpreted by
Hildebrand \& Dragovan (1995) to correspond to lines of sight where the
magnetic field lies in the plane of the sky. The discrepancy between 
$10\%$ polarization observed and the predicted $\sim 35\%$ expected for
the given grain geometry was interpreted by Hildebrand \& Dragovan (1995)
as a consequence of limited efficiency of Gold alignment calculated in
Lazarian (1994). We, however, would tend to believe that the fluctuations
of magnetic field direction are mostly responsible for the discrepancy,
while the alignment is close to being complete (see below). For instance,
polarization
measurements from BIMA that resolve smaller scale structure that those
by Hildebrand's group
 reveal higher degrees of polarization (Goodman 1999).   
In the diffuse gas the structure of magnetic field should be more ordered
and we expect to observe polarization higher than 10\%.

We claim below that the grains responsible for the FIR emission in dark clouds constitute
just a subset of the total grain population, namely, a subset of grains
that are close to the stars and are aligned because of this.
The rest of the grains within dark clouds may not be aligned and
therefore the observed optical and near-infrared polarization does 
not follow extinction for large optical depths (Goodman 1995). 
This reduction of the grain alignment efficiency is nevertheless a characteristic of
high column density regions, and is not likely to appear in the diffuse
ISM. Thus for our modeling of the polarized emission of dust in the 
``infrared cirrus'' at high galactic latitudes (which are more representative
of the foreground emission for CMB polarization measurements), we will
assume that the grain alignment efficiency is perfect.

We should finally wonder about the wavelength dependence of the polarization
degree in emission in the FIR and submillimeter wavelengths. According to
the theory, because the wavelengths considered are much greater than the
maximum size of the grains, this polarization degree should be fairly 
independent of wavelength (see Draine \& Lee 1984). This has been confirmed
by multifrequency observations (ranging from $1300\mu{\rm m}$ to
$100\mu{\rm m}$, see Leach {\em et al.} 1991, Schleuning
{\em et al.} 1996, Hildebrand {\em et al.} 1995). 

\subsection{Grain alignment and polarization}

Grain alignment is a problem of half a century standing with the first
observations dated as far back as 1949 (Hiltner 1949, Hall 1949) and
first theories put forward in 1951 (Spitzer \& Tukey 1951, Davis \&
Greenstein 1951, Gold 1951). Intensive recent research have partially
uncovered the mystery of how grains can be aligned and we briefly summarize
below what is known of grain alignment.

Various mechanism of alignment (see table~1 in Lazarian, Goodman \& Myers 1997,
henceforth LGM97)
can be broadly separated in three categories: {\it paramagnetic}, 
{\it mechanical} and
{\it via radiative torques}. The latter seem to dominate in diffuse
ISM (Draine \& Weingartner 1996, 1997), while mechanical alignment
is important in particular regions of outflows, MHD turbulence
(Lazarian 1994, 1997) and also in the regions of supersonic
ambipolar diffusion (Roberge {\em et al.} 1995). 

The role of paramagnetic effects is not clear so far. Jones and Spitzer
(1967) have shown that grains can be well aligned via paramagnetic
dissipation if they contain ferromagnetic or superparamagnetic inclusions.
This idea was further supported by Mathis (1986), who showed that
such grain can provide a proper wavelength dependence for polarization
and also by Martin (1995) and Goodman \& Whittet (1996) who observed that
GEMs\footnote{A recent study by Draine \& Lazarian (1999a) has shown
that not more than 5\% of iron can be in the form of metal as otherwise
emissivity at 90~GHz would be stronger than observed.} 
(glass with embedded metal) and other strongly magnetic inclusions
found in primitive meteorites and in the interstellar dust particles caught in
Earth atmosphere can render dust the required magnetic response.

Work by Lazarian \& Draine (1997) made the case of paramagnetic alignment
stronger: they showed that grains can be nearly perfectly aligned even
when they have properties of ordinary paramagnetic materials, provided
that grains rotate via H$_2$ torques discovered by Purcell (1979).
However, further research showed that Purcell's torques can be suppressed
by thermal flips of interstellar grains (Lazarian \& Draine 1999 a,b) and
the resulting alignment may be small. The thermal flipping should be
enhanced for grains with superparamagnetic and ferromagnetic inclusions
and therefore these grains are likely to rotate thermally (Lazarian \& Draine
1999a) which entail the decrease of alignment for ``supermagnetic'' grains. 

Radiative torques unlike Purcell torques are not fixed in the grain
body frame. Therefore they are less sensitive to averaging arising from
thermal flipping. Therefore these torques are likely to be the principal
mechanism for grain spin-up and grain alignment in diffuse ISM.
Unfortunately, we do not have proper theoretical understanding of
the alignment via radiative torques. The existing numerical simulations
disregard grain flipping and therefore cannot provide the quantitative
answer. However, it seems plausible that the degree of 
radiative alignment should be high, e.g. of the order $\sim 80$\%.

The picture of grain alignment in diffuse media is simpler than in 
molecular clouds, where grains in cold dense environments can come
to thermal equilibrium with gas and randomize.
According to LGM97 the latter process is
responsible for marginal polarizing power of grains in dark clouds
observed by Goodman (1995). Recently Hildebrand {\em et al.} (1999) confirmed
this prediction of LGM97, observing that cold grains are not aligned,
while hot grains are aligned.

\section{Angular distribution of polarized dust emission}

\subsection{The problem}

Because the grains align themselves on the magnetic field, and
polarize the light (absorption or emission) by dichroism, it is
clearly tempting to try to infer the distribution of the underlying
magnetic field from polarization maps. The problem is however far
from being trivial. As we saw, the ratio of polarization to extinction
is sensitive to a large number of factors, which include the random changes
of the turbulent part of the magnetic field, as well as possible changes 
of the polarization efficiency of grains along the line of sight, and
from one line of sight to the other. However, observations seem to
show in a robust way that the polarization efficiency is decreased only
in the denser parts of the interstellar objects (where $A_V\gg 1$), and
is maximal in the diffuse interstellar medium (LGM97),
so we will concentrate mainly on the distribution of magnetic field flux.

The problem has been first studied by Chandrasekhar \& Fermi (1953) in
the case of wave-like fluctuations of the magnetic field (describing the
case of Alfvenic turbulence). They used the distribution of polarization
position angles to infer the amplitude of the mean magnetic field. 
This approach was further elaborated by Zweibel (1990) in the case of
a collection of clouds threaded on the magnetic field lines. The value
of the magnetic field amplitude computed by this method is in good 
agreement with values inferred from completely different methods, like
Zeeman splitting of the HI line (see Heiles 1987), and provides additional
indication that the polarization patterns are indeed related to the 
underlying magnetic field distribution. 

Jokipii \& Parker (1969) addressed the statistics of magnetic field 
in a more general way. They concluded that in order
for cosmic rays to escape the Galactic disk in about $1 {\rm Myr}$, the
correlation length of the magnetic field lines had to be of the order
of $100 {\rm pc}$ in the diffuse ISM. 
This method has been extended by Myers \& Goodman (1990), who inferred from
the observation of the distribution of polarization position angles in
dark clouds the relative amplitudes of
the random and regular components of the magnetic field, as well as the 
amplitude and direction of the mean field (with the addition of Zeeman
splitting measurements in the same regions, see Heiles 1987,1988). 
They obtained a rough estimate for the number of
magnetic field correlation lengths along the line of sight. 

A unifying approach was adopted by Jones, Klebbe \& Dickey
(1992). They describe the
magnetic field fluctuations in a statistical way, but relate its correlation
length to the optical depth variations along the line of sight, thereby
stressing the coupling between the magnetic field and the underlying density
field. They generalized the earlier studies by the introduction of a 
correlation function of the magnetic field position angle along the line
of sight, as a function of optical depth. They obtained an estimate of
the typical correlation length of the magnetic field of the order of 
$\Delta\tau^K_0 = 0.1$.

The idea of relating the magnetic field directions
distribution to the underlying density field seems physically
well motivated.
Unfortunately, it is difficult to quantify the correlation between
interstellar density and magnetic field. Some insight could be obtained
via numerical simulations of compressible MHD. The results available
thus far deal unfortunately either with 2D turbulence
(Vazquez-Semadeni, Passot \& Pouquet 1995; Passot, Vazquez-Semadeni \& 
Pouquet 1995; Vazquez-Semadeni, Passot \& Pouquet 1996) or with 3D turbulence
(Padoan \& Nordlund 1998, Jones, Ostriker \& Gammie 1998),
but with insufficient resolution to obtain a detailed 
statistical picture. 
These simulations, however, do indicate 
a correlation of magnetic field and the underlying gas density.

The kinds of statistical approaches described above may be useful
in addressing the problem of predicting the polarized foreground from dust in
the diffuse ISM since the CMB will also be analyzed
in term of its statistical properties. To be more practical, we
would like to compare not only the one-point PDF of both emissions,
but also their two-point functions, or equivalently their spatial
power spectra. This problem, to be solved, requires not
only a knowledge of the dust density and magnetic field distribution,
but also the knowledge of their {\em cross-correlation} which
seems to be non-vanishing, as shown by numerical simulations.   

Unfortunately, only some derived statistical properties (like projected
distributions) are available through the observations. A fair
amount of modeling will then be necessary to guess what the power
spectra of the dust polarized emission could look like. We will 
present different possible ways to estimate its properties in the following.

\subsection{Getting some ideas about the dust polarized power spectra}

\begin{figure}
\plotfiddle{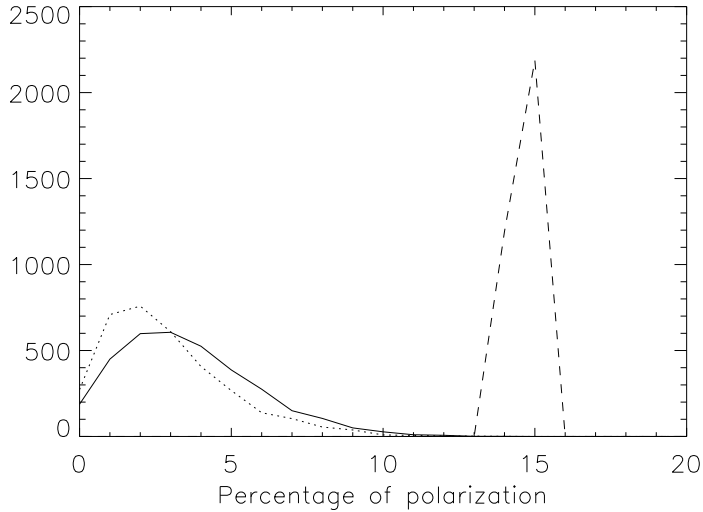}{6cm}{0}{60}{60}{-176}{-230}
\caption{Histogram of the polarization degree of dust emission in three
different cases for the correlation magnetic field-density. {\em Solid lines} 
correspond to the case where field lines are aligned with the gas filaments,
{\em dotted lines} correspond to random field lines perpendicular to the 
filaments, and {\em dashed lines} to a constant magnetic field with an angle
$\zeta = \pi/2$ from the plane of the sky. After PSBM98. \label{fig:percentage}}
\end{figure}

A first (and rather crude) model to infer the statistical properties
of dust polarized emission has been discussed in Prunet {\em et al.} (1998, henceforth
PSBM98). 
The fundamental ideas behind their method are the following:
\begin{itemize}
\item[$\bullet$] The dust emissivity (as seen by IRAS/DIRBE maps at $100
\mu{\rm m}$) is strongly correlated to the underlying gas density
(as seen by the HI $21 {\rm cm}$ line maps of the Galaxy, see Hartmann \&
Burton 1995), at least for column densities smaller than $5\times 10^{20}
{\rm cm}^{-2}$ (Boulanger {\em et al.} 1996). 
\item[$\bullet$] The HI maps contain an additional information, which is
the distribution of HI densities in velocity space along each line of sight
(measured through the Doppler shift of the HI emission lines), and therefore
contain some information about the statistical properties of the 3D 
distribution of HI gas. 
\end{itemize}
PSBM98 used the galaxy rotation curve to map the HI density field from velocity
space to real space, and then assumed three different possibilities for
the correlation of the magnetic field lines distribution with respect to
the underlying HI density distribution:
\begin{itemize}
\item[1.] The magnetic field lines follow the filamentary HI structures
(i.e. are locally aligned with the density least-gradient direction).
\item[2.] The magnetic field lines are randomly chosen in the plane 
perpendicular to the local direction of the filaments (this could be 
representative of helicoidal field lines wrapped around filaments).
\item[3.] The magnetic field is constant throughout the observed patch
(as a comparative case)
\end{itemize}
Those prescriptions, together with the arguments provided in 
sections~\ref{sec:shapes} and \ref{sec:emissivity} allowed PSBM98
to infer the spatial power spectra of the Stokes parameters of
the dust polarized emission, as well as their cross-correlation power
spectra with unpolarized emission (hereafter referred to as temperature 
emission). However, the first observable quantity predicted by this method 
is the distribution of polarization degree in emission, which can be compared
to the measured histograms of the same quantity toward dense objects
(see Hildebrand 1996 and references therein). This histogram is shown in
Fig.~\ref{fig:percentage}. 
We can see immediately that the uniform magnetic field case does not 
reproduce the observations (as expected), but the first two cases seem to 
reproduce correctly the observational trend, with a peak at a few percent and a long
tail extending up to $\sim 12\%$. However, it was argued in section 
\ref{sec:emissivity} that an overall reduced 
polarization (by a factor of $\sim 4$) was 
necessary to explain the 
observed polarization degrees in dark clouds. 

The histograms shown in
Fig.~\ref{fig:percentage} are supposedly representative of the
diffuse ISM where the grain alignment is expected to be nearly perfect. 
Although they peak at slightly higher values of 
polarization (in the two realistic cases) compared to the histograms observed in dark clouds, 
they can not reproduce this factor of 
$4$. This discrepancy may have numerous reasons. First, this value of $4$ for
the reduction of polarization in dark clouds, was interpreted by Hildebrand \& Dragovan 
(1995) entirely as a reduction of the grain alignment efficiency, under the assumption 
that the maximum observed polarization level was representative of a 
{\em single} magnetic field line in the plane of the sky; or if the field had 
already experienced one or two reversals along the line of sight, this factor
would have to be lowered. Another possible explanation comes from the naive
mapping used in the model to go from velocity space to real space. Indeed, the presence
of the turbulent velocities can artificially stretch the structures along
the line of sight, thereby artificially reducing the amount of polarization in the case 
where magnetic field lines are aligned with the filaments. 

Finally, the 
(almost) random magnetic field case is giving too low results, as expected
since it does not reproduce the smooth magnetic field component observed
in the galactic disc (called ``uniform'' or ``regular'' magnetic field). The latter is
necessary to reproduce the observed polarization to extinction ratios in the
near infra-red (see Jones, Klebbe \& Dickey 1992), as well as the position
angles distribution (see Myers \& Goodman 1991). 
We can then consider the constant magnetic field case and the random magnetic
field case as {\em limiting cases} for this study. We will see in what
follows that they should give some upper and lower bounds on the 
{\em slope} of the power spectrum of polarization, as well as a rough 
normalization.   

\begin{figure}
\plottwo{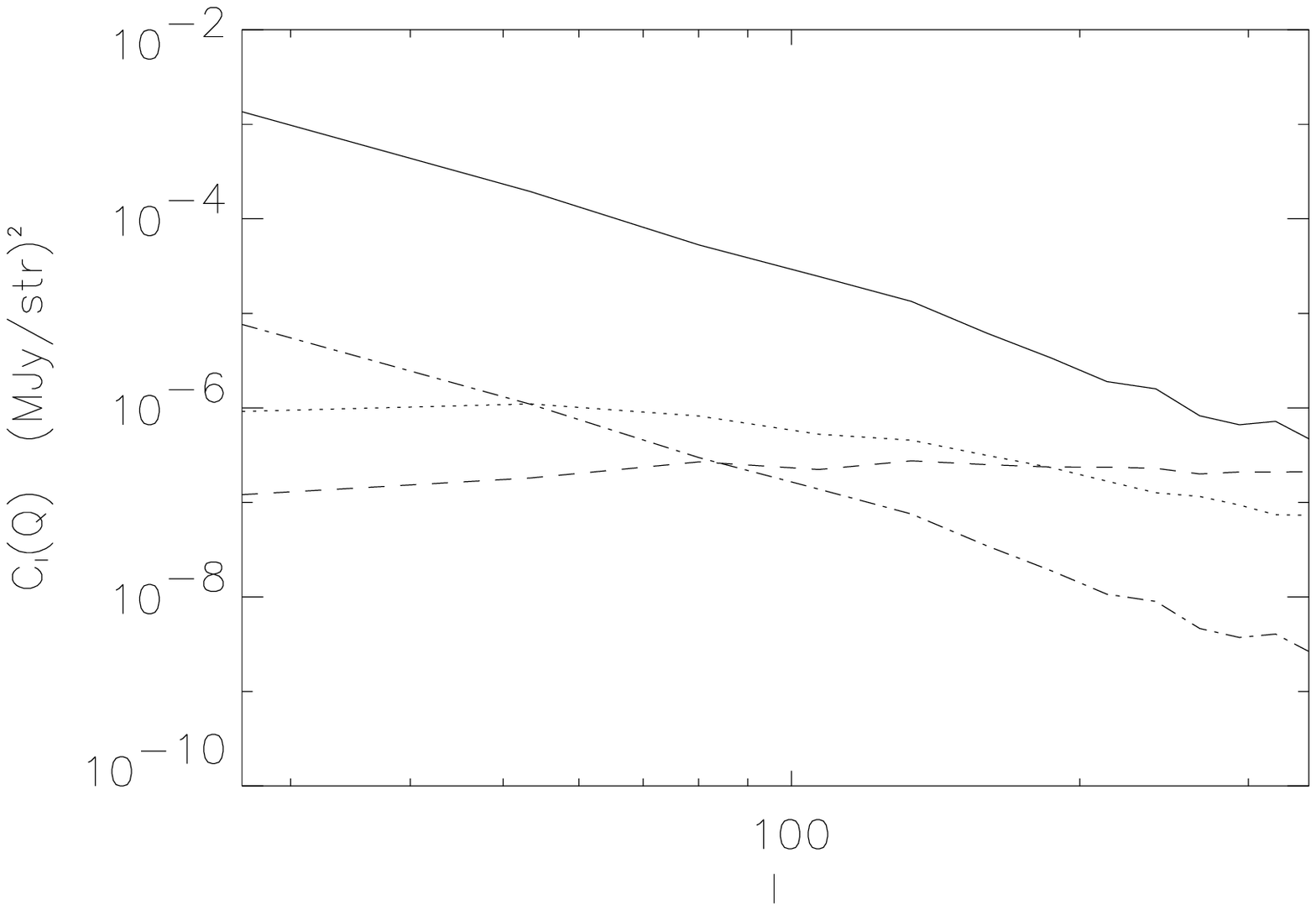}{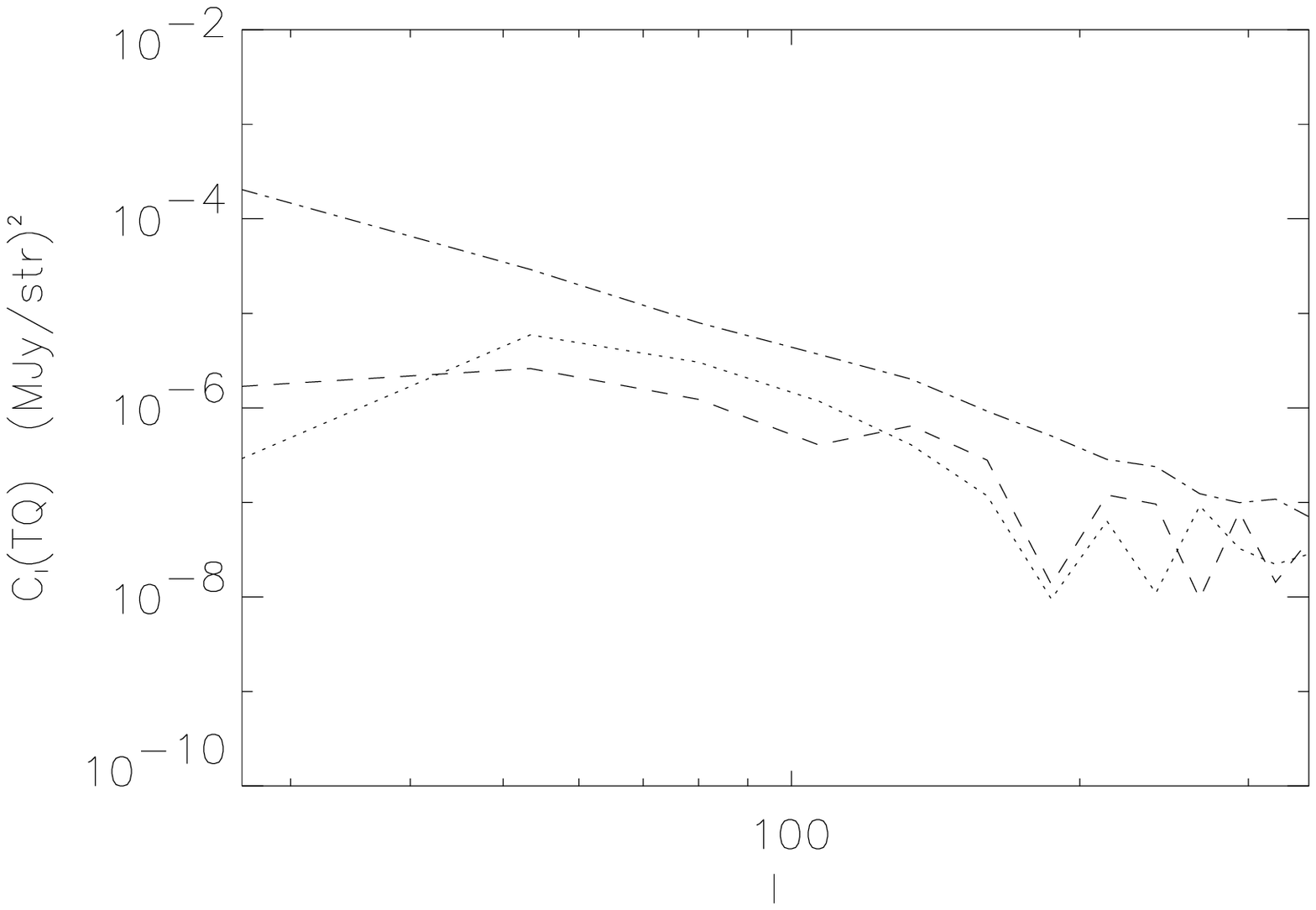}
\caption{ {\em Left Panel:} Power spectra of the Stokes parameter Q in the 
three cases for
the magnetic field: {\em dotted line} for field lines along gas structures,
{\em dashed line} for random field lines perpendicular to gas structures, 
and {\em dot-dashed line} for the constant field case. The power spectrum 
of the unpolarized emission is shown for comparison ({\em solid line}).
{\em Right panel:} Power spectra of the cross-correlation between the 
temperature and Stokes parameter Q, same line styles.
Figure after PSBM98.\label{fig:stokes} }
\end{figure}

The power spectra of the Stokes parameter Q are shown in Fig.~\ref{fig:stokes}
for the three different assumptions for the magnetic field lines. 
The case with constant magnetic field has, as expected, the same power-law
dependence as the unpolarized emission ($C(\ell)\propto \ell^{-3}$, see
Gautier {\em et al.} 1992). What is more interesting is the fact that the
two other power spectra are both flatter; indeed, even if the correlation
length of the magnetic field is similar to the density's one, the {\em 
modulation} of the density field by the magnetic field orientation results
always in a flatter power spectrum. Then we can conclude that the slope 
of the real power spectrum of the polarized dust emission should lie
somewhere between those given by the constant and random magnetic field cases.

Another quantity of interest for the measurement of polarization in the 
CMB is the cross-correlation between the polarized and unpolarized emission
(e.g. Zaldarriaga \& Seljak 1997). The cross-correlation power spectra 
are given in Fig.~\ref{fig:stokes} for the dust, and for the three assumptions
concerning the magnetic field. We can see from this figure that this spectrum
is expected to have more or less the same value around the degree scale,
which is of interest for CMB polarization measurements (see next section). 
For a mathematical definition of the power spectra, see PSBM98.

This model has several advantages and drawbacks. The main advantage of this
model is to put upper and lower limits on the slope and normalization of
the different polarized power spectra expected for diffuse dust emission,
which will be of interest for a comparison to the predicted CMB polarized
power spectra. From this point of view it gives a rough first idea of what
we should be dealing with when processing CMB polarized data.
However, it does not give a precise picture of what should be the polarization
patterns of dust emission in any localized region. Moreover, it could be
that the predicted statistical observables themselves may be flawed by 
the presence of the turbulent motions that spoil any attempt to do a precise
mapping of HI density from velocity space to real space. It appears that only
in the case of short wavelength dominated density spectra can we hope
to measure the statistical properties of the gas density field from 
HI data (see Lazarian \& Pogosyan 1999).
We will then give some guidelines for alternative methods to derive the
polarized spectra of dust emission, taking into account the turbulent
velocity component of HI, or relying on 2D data only.

\section{Comparison with CMB power spectra}

\begin{figure}
\plotfiddle{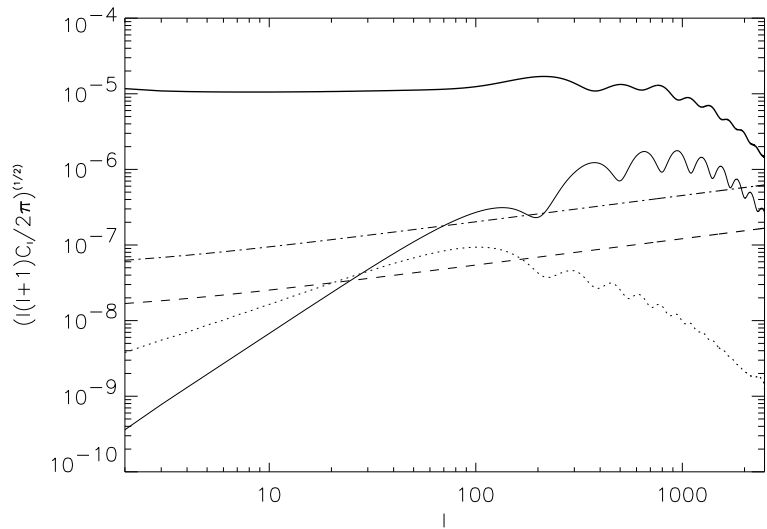}{6cm}{0}{60}{60}{-180}{-220}
\caption{The power spectra of the E modes of polarization are shown for dust
at $143\,{\rm GHz}$ ({\em dashed line}) and $217\,{\rm GHz}$ ({\em dot-dashed line}).
The scalar-induced ({\em solid line}) and tensor-induced ({\em dotted line})
power spectra of CMB E-modes are also shown. The CMB temperature power spectrum 
({\em thick solid line}) is shown for comparison. $143$ and $217\,{\rm GHz}$ are
the central frequencies of the two most sensitive polarized channels of \sc{planck}.
\label{fig:E}}
\end{figure}

\begin{figure}
\plotfiddle{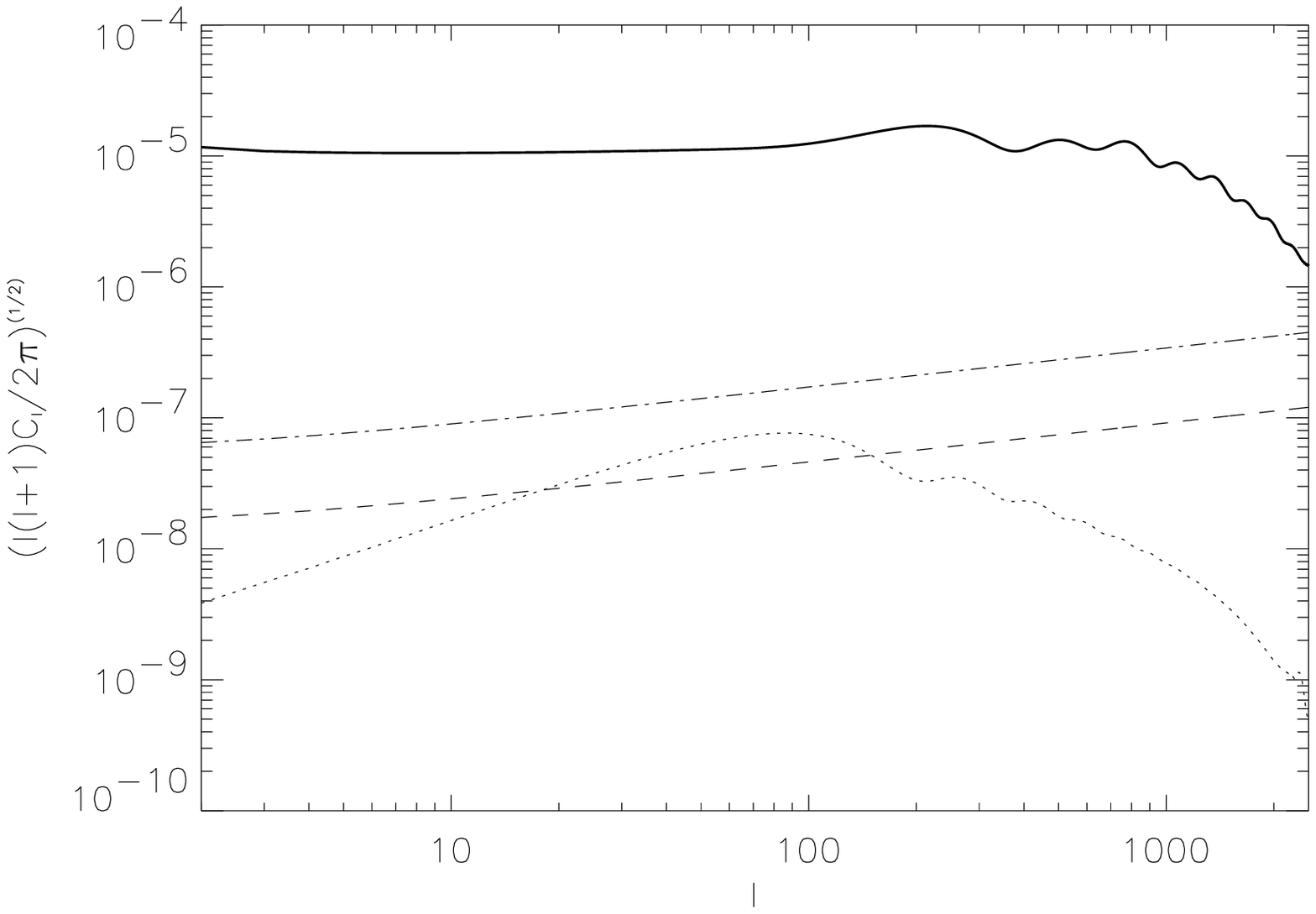}{5cm}{0}{60}{60}{-177}{-220}
\caption{Same as Fig.~\ref{fig:E}, but for the B modes of polarization.
\label{fig:B}}
\end{figure}

\begin{figure}
\plotfiddle{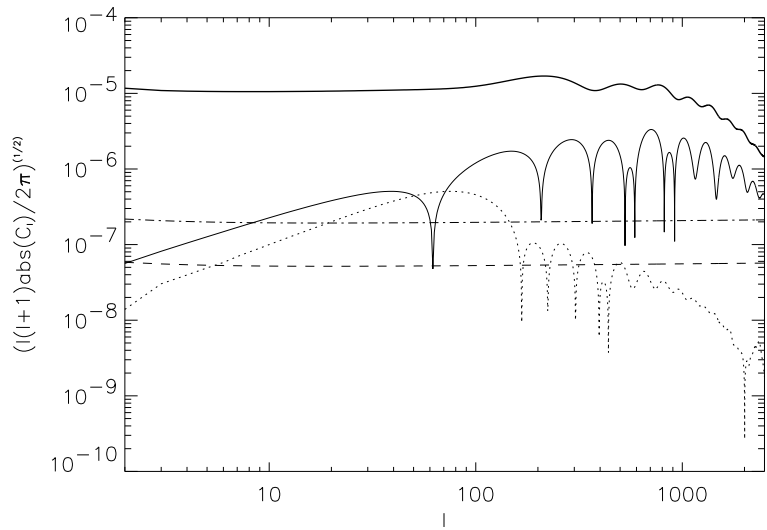}{7cm}{0}{60}{60}{-177}{-220}
\caption{Same as Fig.~\ref{fig:E}, but for the T-E correlation.
\label{fig:TE}}
\end{figure}

Finally, we would like to compare the amplitude of the different polarized
power spectra of the dust emission from the diffuse ISM to the {\em predicted}
amplitudes of the same quantities for the CMB, as inferred by theoretical
computations (Seljak \& Zaldarriaga 1996). It is most convenient to redefine
new variables of polarization (Zaldarriaga \& Seljak 1997), called ``electric''
and ``magnetic'' modes of polarization ($E,B$) in analogy with the transformation
properties of the electro-magnetic field under parity transformations. These modes
are particularly well adapted to describe CMB polarization, as scalar perturbations
induce only electric modes of polarization (Seljak \& Zaldarriaga 1997, see also
Kamionkowski {\em et al.} 1997a,b for a similar analysis in terms of tensor
spherical harmonics). As we dealt with relatively small patches on the sky in
the analysis performed on the HI maps (see PSBM98), we can use
the definition of those modes in the small-angle (or flat space) approximation 
(Seljak 1997): 
\begin{eqnarray}
C_E(\ell) &=& {1\over N}\sum_1^N \left| Q(\vec{\ell})\cos(2\phi_{\vec{\ell}})
+ U(\vec{\ell})\sin(2\phi_{\vec{\ell}}) \right|^2 \\
C_B(\ell) &=& {1\over N}\sum_1^N \left| -Q(\vec{\ell})\sin(2\phi_{\vec{\ell}})
+ U(\vec{\ell})\cos(2\phi_{\vec{\ell}}) \right|^2 \\
C_{TE}(\ell) &=& {1\over N}\sum_1^N \left[ \left( Q(\vec{\ell})T^*(\vec{\ell})
+ Q^*(\vec{\ell})T(\vec{\ell}) \right)\cos(2\phi_{\vec{\ell}}) \right. \nonumber\\ 
&&+ \left.
\left( U(\vec{\ell})T^*(\vec{\ell}) +U^*(\vec{\ell})T(\vec{\ell}) \right)\sin(2\phi_{\vec{\ell}})
 \right]
\end{eqnarray}
The different power spectra for CMB and dust are shown in figures~\ref{fig:E},\ref{fig:B}
and \ref{fig:TE}
in the case where the magnetic field lines are assumed to follow the directions
of the gaseous filaments. The CMB power spectra have been computed for a tilted 
CDM model with scalar spectral index $n_S = 0.9$, tensor index $n_T = n_S -1$ and
a tensor to scalar temperature quadrupole ratio of $T/S=7n_T$, using the Boltzmann
code CMBFAST (Seljak \& Zaldarriaga 1996). 
We can draw some interesting conclusions from those figures. If we keep
in mind the uncertainties discussed before concerning the dust polarized 
power spectra, we can conclude that the E modes power spectra of CMB should
be measurable at least for scales smaller than the degree ({\em i.e.} in the
Doppler peaks region). Concerning the temperature-polarization cross-correlation,
we expect, as shown in figure~\ref{fig:TE}, that the correlation will be stronger
for the CMB signal than for the dust signal, making the CMB signal
accessible to measurements up to even larger scales ($\sim 5^\circ$). 
However, the B-modes of CMB polarization are {\em of the same level}
as their contaminants, and we cannot at this stage conclude about the
possibility of their measurement. 

\begin{figure}
\plotfiddle{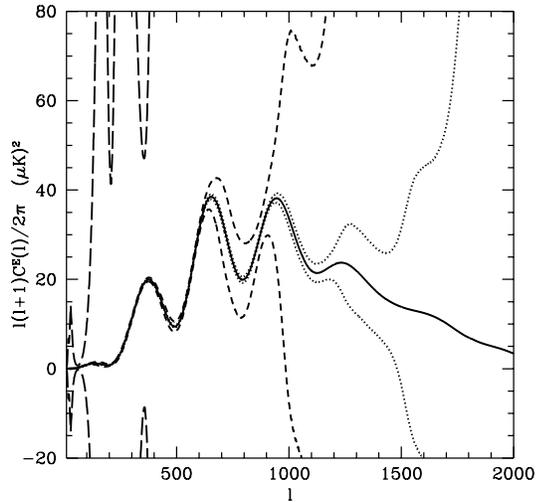}{8cm}{0}{35}{35}{-140}{-40}
\caption{Band-power estimates of the $1-\sigma$ error on the measurement of the
E-modes power spectrum, for the {\sc planck}-HFI ({\em dotted line}), 
{\sc planck}-LFI ({\em short-dashed line}) and MAP ({\em long-dashed line})
configurations. The bands have constant logarithmic width $\Delta\ell/\ell=0.2$.
\label{fig:bandE}}
\end{figure}

\begin{figure}
\plotfiddle{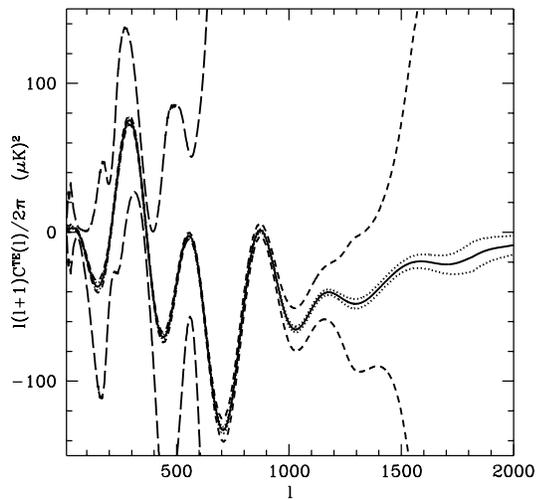}{7.5cm}{0}{35}{35}{-140}{-40}
\caption{Same as Fig.~\ref{fig:bandE}, but for the T-E cross-correlation.
\label{fig:bandTE}}
\end{figure}

\begin{figure}
\plotfiddle{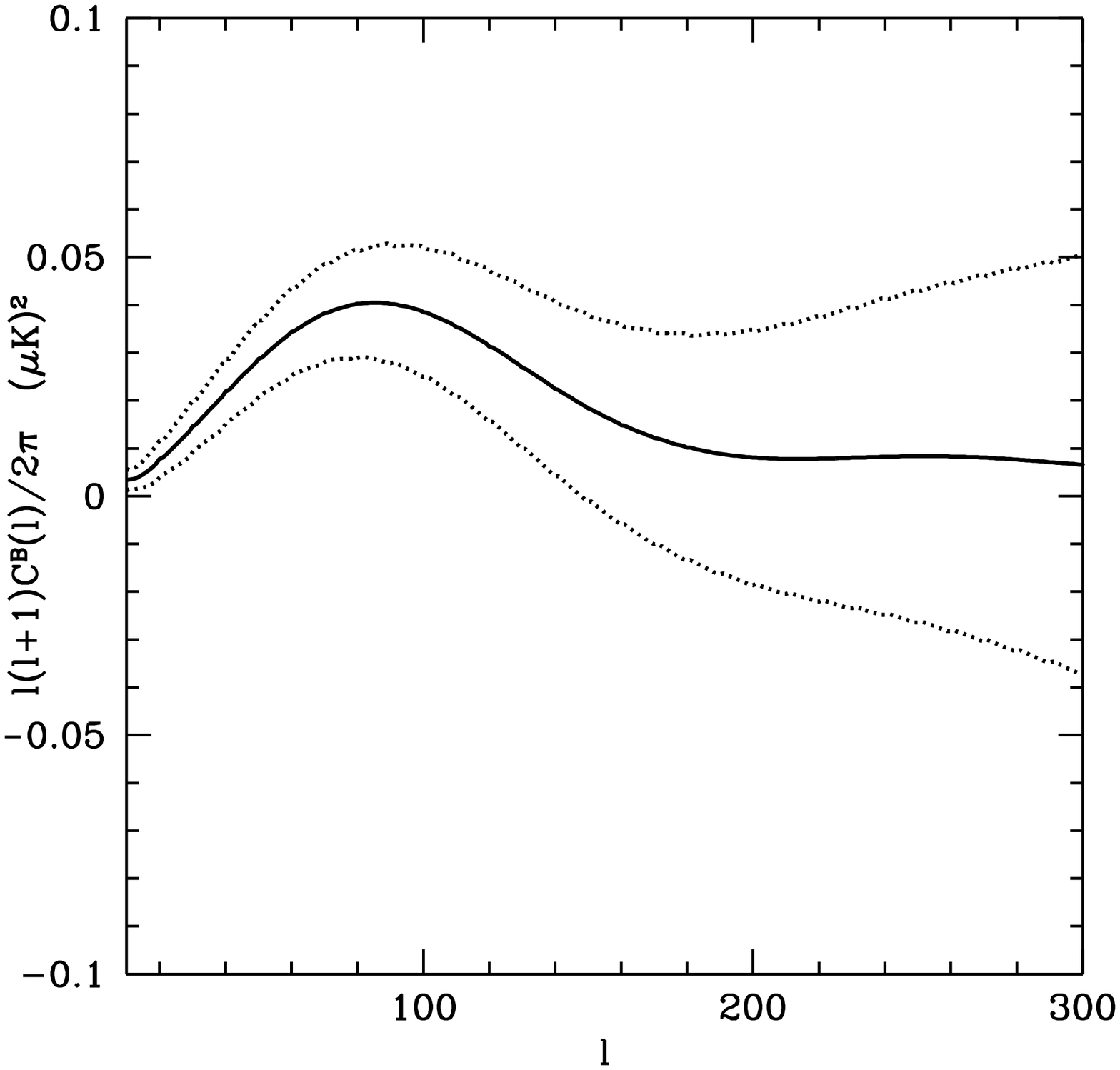}{7.5cm}{0}{35}{35}{-140}{-40}
\caption{Same as Fig.~\ref{fig:bandE}, but for the B-modes of polarization.
\label{fig:bandB}}
\end{figure}

We discussed earlier the fact that the degree of polarization of
the dust {\em thermal} emission should be fairly independent of
wavelengths. Then, if we learn about the spectral dependence of
dust emissivity on wavelength by observing denser regions of
the ISM, we should be able to predict with a good accuracy the
spectral behavior of the polarized dust from the diffuse ISM
(this is subject to the fact that the denser regions observed
to measure the emissivity of dust at different wavelengths 
are close in composition and grain temperatures to the very
diffuse ISM; we therefore refer to the former as {\em intermediate
density regions}.) We can then use this information, combined with
the polarized maps observed at higher frequencies where the dust
emission is {\em dominant} over all other processes (including CMB), 
to {\em clean} the low frequency polarized maps where we seek the
CMB signal. This multi-frequency filtering of the data can be implemented
in a lot of different ways, and has been applied to CMB anisotropy data
by Kogut {\em et al.} (1996) for the first time on the COBE DMR data.
Since then, different techniques have been developed, which can be grouped
in two broad categories: Wiener filtering techniques (Bouchet {\em et al.} 1996,
Tegmark \& Efstathiou 1996),
and Maximum Entropy Methods (Maisinger {\em et al.} 1997, 
Hobson {\em et al.} 1998). Bouchet {\em et al.} (1999)
generalized the Wiener filtering method to take into account 
polarized CMB data. The resulting errors on the polarized power spectra,
{\em after removal} of the foreground emissions (which were in this
article Galactic dust and synchrotron polarized emissions) are shown
in figures~\ref{fig:bandE}, \ref{fig:bandTE} and \ref{fig:bandB}
for different instrumental configurations.  
We have to be careful in reading those figures, as a logarithmic smoothing
has been performed; the covariances shown here are for band-power estimates
of the power spectra, not for individual modes. They give nevertheless a good
idea of what accuracy should be reached by the upcoming satellite missions 
in the observation of CMB polarization. For instance, relatively broad band
features in the spectra of E modes and cross-correlation should be observed
with a good accuracy. The B modes power spectrum is, as expected from the
weakness of the signal, much less constrained. However, even a poor determination
of this signal is important for constraining tensor-induced cosmological
parameters (Prunet {\em et al.} 1998a,1998b). We should also notice that
the amount of gravitational waves could be much higher for a given class
of inflationary models (Lesgourgues {\em et al.} 1998), thus making the signal
of B modes easier to measure. 

\section{Discussion}

\subsection{Various approaches}

Out results sensitively depend on the assumed angular distribution of the
polarized dust emission. The model used so far has its flaws. It is
therefore important to test other approaches. 

\paragraph{A full statistical approach:}

As shown by Lazarian \& Pogosyan (1999), the gas underlying density
distribution in the diffuse ISM should be measurable in the HI velocity
space data cubes (Hartmann \& Burton 1995) if it is short-wave dominated.
With some assumption about the correlation of the field lines with the
density, inspired by numerical simulations and/or theoretical considerations
of MHD turbulence, one could infer the resulting angular distribution
of polarized dust assuming perfect, or non perfect alignment. This
method, under the above assumption concerning the magnetic field-density
correlation, would at least be correct statistically, which is the primary
goal of the exercise. The measurement of polarization statistics by upcoming
satellite missions would then
be a mean to check the validity of theories of MHD compressible turbulence !

\paragraph{A pragmatic approach:} 

Whether or not the underlying statistics of HI is short- or long-wave
dominated the density distribution integrated over 21~cm emission line   
reveal HI density structure. To find the statistics of the 3D 
underlying density one may use the inversion technique in Lazarian
(1995) or use the observe 2D structure to infer the expected projected
pattern of magnetic field. If magnetic field direction fluctuates along
the line of sight this method should provide an overestimate of
the expected variations of polarization.

\paragraph{Using synchrotron:}

Synchrotron emissivity depends both on the distribution of magnetic 
field and the distribution of relativistic cosmic electrons. The 
latter are distributed more smoothly than the former and this allows to
infer the small-scale statistics from variations of synchrotron
intensity\footnote{Variations of synchrotron polarization is another
channel of valuable information.} (see Lazarian 1992). 
Potentially synchrotron is a very valuable
tool for studying the statistics of magnetic fields. Having this
statistics and 2D statistics of HI emissivity it seems possible to
evaluate the expected polarized emissivity.

\subsection{Polarization at lower frequencies}

So far we have discussed the polarization at frequencies larger than
100~GHz. At lower frequencies dust contributes to microwave emissivity
via magneto-dipole emission and rotational emission from ultra-small
grains (Draine \& Lazarian 1999b)and both types of emission are partially
polarized. The magneto-dipole emission is most prominent when the
grain materials are strongly magnetic, e.g. ferro- or ferrimagnetic\footnote{
With iron and nickel constituting up to 30 percent of ``silicate'' grains,
these properties do not look unrealistic.}. However, even for
ordinary paramagnetic grains magneto-dipole emission dominates the 
ordinary electro-dipole emission for frequencies less than 30~GHz.
The emission from single dipole ferro or ferrimagnetic 
grains is strongly polarized (up to 40\%) and shows characteristic
frequency dependence signatures (Draine \& Lazarian 1999b).

Emission from ultra-small grains emanates due to grain rotation.
Small grains are likely to have dipole moment and therefore the emission
can be called rotational emission in analogy with the corresponding
emission from molecular transitions. If small grains are aligned
their emission is polarized. Calculations in Lazarian \& Draine (1999c)
show that polarization degrees up to a few percent is possible for
rotational emission at frequencies less than 40~GHz.

\section{Conclusions}

Our results can be summarized as follows:

1. Polarized radiation from dust can appreciably interfere with the attempts
to measure polarization of cosmological origin. The signal from tensor
modes is expected to be completely masked by dust. However, with the expected 
independence of the dust polarization degree on wavelength, we can hope 
to remove this contaminant from CMB maps with reasonable efficiency, provided that we
have polarized measurements of the same regions at higher frequencies where the
dust emission is dominant.

2. The degree of dust polarization is expected to vary with scale with the
tangling of magnetic field limiting the achievable polarization.

3. The expected polarization from dust differs in the range of high
($>100$~GHz) and low ($<100$~GHz) frequencies. The emission at high
frequencies is of electro-dipole nature and therefore can be treated
using the accepted technique. The low frequency emission is either of
magneto-dipole origin or from rotating ultra-small grains and the
expected polarization depends on various factors, e.g. on whether
grains consist of single magnetic domains or have multi-domain 
structure.

4. Multi-frequency modeling of the galactic polarization on the basis of
synchrotron, HI and infrared polarization maps should allow 
high-precision study of cosmological polarization.

\end{document}